\documentclass[pdflatex,sn-mathphys-num]{sn-jnl}

\usepackage{graphicx}%
\usepackage{multirow}%
\usepackage{amsmath,amssymb,amsfonts}%
\usepackage{amsthm}%
\usepackage{mathrsfs}%
\usepackage[title]{appendix}%
\usepackage{xcolor}%
\usepackage{textcomp}%
\usepackage{manyfoot}%
\usepackage{booktabs}%
\usepackage{algorithm}%
\usepackage{algorithmicx}%
\usepackage{algpseudocode}%
\usepackage{listings}%
\usepackage{bbold}
\usepackage{braket}

\newcommand{\oprod}[2]{| #1 \rangle\!\langle #2 |}

\newcommand{\id}{\mathbb{1}}
\DeclareMathOperator{\Tr}{Tr}
\DeclareMathOperator{\tr}{Tr}

\begin{document}

\title[Quantum mereology and subsystems from the spectrum]{Quantum mereology and subsystems from the spectrum}


\author*[1]{\fnm{Nicolas} \sur{Loizeau}}\email{nicolas.loizeau@nbi.ku.dk}

\author[2,3]{\fnm{Dries} \sur{Sels}}

\affil[1]{\orgdiv{Niels Bohr Institute}, \orgname{University of Copenhagen}, \orgaddress{\city{Copenhagen}, \country{Denmark}}}

\affil[2]{\orgdiv{Department of Physics}, \orgname{New York University}, \orgaddress{\city{New York},  \state{NY}, \country{USA}}}

\affil[3]{\orgdiv{Center for Computational Quantum Physics}, \orgname{Flatiron Institute}, \orgaddress{\city{New York},  \state{NY}, \country{USA}}}


\abstract{
The minimal ingredients to describe a quantum system are a Hamiltonian, an initial state, and a preferred tensor product structure that encodes a decomposition into subsystems. We explore a top-down approach in which the subsystems emerge from the spectrum of the whole system. This approach has been referred to as quantum mereology.
First we show that decomposing a system into subsystems is equivalent to decomposing a spectrum into other spectra.
Then we argue that the number of subsystems (the volume of the system) can be inferred from the spectrum itself. In local models, this information is encoded in finite size corrections to the Gaussian density of states.}




\maketitle

\section{Introduction}

The idea of the atom emerged when Democritus asked: Can one divide a physical object into two parts indefinitely? It originated from a top-down approach. In contrast, most of modern physics relies on a reductionist bottom-up approach: the existence of particles is presupposed as an axiom of the theory, and macroscopic objects emerge from the interactions between particles.

In this paper, we explore the top-down approach, proposing that particles emerge from a holistic description of the world, rather than building the world out of subsystems.

Consider an extremist point of view where the world is reducible to a finite-dimensional quantum mechanical system \cite{Carroll2019, Carroll2021, Carroll2022}. Then, the world is fully characterized by two objects: its Hamiltonian $H$ and its initial state $\psi_0$. In the eigenenergy basis, these two objects can be expressed as two sets of numbers: $E_n$ the eigenvalues of $H$ and $c_n$ the entries of $\psi_0$ in the eigenbasis of $H$. Since quantum mechanics is independent of the choice of basis, one could consider these two sets of numbers to be the most fundamental pieces of information that describe the world. However, to identify this representation with our empirical experience, we need to introduce an additional structure \cite{Sels2014}. This structure can take the form of a preferred set of observables or a preferred basis that encodes a particular tensor product structure. This particular tensor product structure yields a decomposition of the Hilbert space into subsystems. For example, when one writes a quantum Hamiltonian, e.g $H=\sum_i Z_iZ_{i+1}+\sum_i X_i$ one writes it in a preferred basis encoding a preferred tensor product structure that is induced by the observables we are interested in measuring, (e.g a decomposition into spins) \cite{Zanardi2004}. Different tensor product structures are related by a unitary transformation. For example, the spin system from above can be mapped to a spinless fermion model through the Jordan-Wigner transformation. In this example, the Jordan-Wigner transformation relates two different tensor product structures that encode different decompositions into subsystems : spins or fermions.

Now, if we convince ourselves that the spectrum and initial state are the most fundamental objects, we need to suggest a mechanism that explains the emergence of preferred tensor structures, i.e., explain why we see the world as a set of hierarchically organized subsystems. This approach is referred to as quantum mereology and has gathered recent interest \cite{Carroll2019, Carroll2021, Carroll2022, Zanardi2024, Adil2024, Cotler2019, Friedrich2024, Cao2017}.

To clarify, let us reformulate the quantum mereology problem as the question: What are the minimal ingredients necessary to describe the framework of quantum mechanics?
In order to do quantum mechanics, one needs a Hilbert space, a Hamiltonian, a state, and a set of preferred observables. The only basis-invariant quantity of a Hamiltonian is its spectrum; therefore, specifying a Hamiltonian without a set of preferred observables is equivalent to specifying only a spectrum. From a mereological point of view, this is the whole: a system taken as a whole, with no information about the parts is just the spectrum of the system. Now comes the question: what are the minimal ingredients required to specify a quantum system? 
In this work, we discuss recovering the preferred observables from the spectrum only. We can indeed argue that the state can be omitted from the fundamental ingredients by introducing the assumption that typical states are low-energy states close to the ground state. In this scenario, the state will inherit its entanglement structure from the structure of the interactions in the Hamiltonian.

Our work aligns with recent advancements which we outline now to provide some context. We share very similar motivations and approaches with Carroll and Singh's work \cite{Carroll2021} which describes a method to factor the Hilbert space into system and environment given a Hamiltonian. Their approach involves searching for a partitioning that yields quasiclassical dynamics. Here, `quasiclassical' refers to the system admitting pointer observables a la Zurek \cite{Zurek1993, Zurek2003} that do not entangle with the environment (or at least entangle as slowly as possible) and remain localized close to classical trajectories. The main limitation of this approach is that the numerics are restricted to very small systems (dimension $\sim 30$) while our simplified approach can deal with actual many-body systems, e.g. order 20 spins (dimension $2^{20}$). Our approach, in fact, leads to similar results in terms of entanglement growth. We will show that by simply minimizing the interaction between two sectors, we can find partitionings that exhibit slow entanglement growth. We do not need to time-evolve any state, in contrast with ref \cite{Carroll2021}.

In ref \cite{Zanardi2022}, Zanardi et al. propose a more generic approach that drops the system-environment structure and aims to partition a system into subsystems that `maintain their informational identities the longest', meaning that information scrambling between subsystems is minimal. They define a scrambling rate that, when minimized, gives rise to emergent subsystems and argue that minimizing the scrambling rate is equivalent to minimizing the interaction strength between subsystems. Our approach employs this idea to further simplify quantum mereology by focusing solely on minimizing the interaction strength between subsystems. Moreover, we will show that the only relevant objects needed to solve this problem are the spectrum of the full system and the spectra of the subsystems. This not only makes our approach numerically more practical, but also philosophically closer to the initial motivations of quantum mereology as stated in ref \cite{Carroll2019}: recovering the subsystems from the spectrum of the world.

Another noteworthy approach to quantum mereology has been proposed by Freedman and Zini \cite{Freedman2021a, Freedman2021b, Zini2023}, who suggest a mechanism that selects for preferred tensor structures and gives rise to subsystems via spontaneous symmetry breaking at the level of the probability distribution of Hamiltonians.

The current work is the natural continuation of a program we started in ref. \cite{l2023} and employs a very similar method. In practice, we restrict the mereology problem to asking: given a spectrum and a particular tensor structure, are they compatible? i.e. is there a Hamiltonian compatible with this tensor structure that has this spectrum? We previously showed that random matrix spectra can be reproduced by 2-local Hamiltonians with exponential precision in system size. In doing so, we showed how a preferred decomposition into qubits can emerge solely from a spectrum. We also showed that these 2-local solutions are not fine-tuned in the sense that they are stable under perturbations of the couplings of the 2-local Hamiltonian. In fact, the irrelevance of a large class of perturbations allowed us to further reduce the number of coupling constants in the Hamiltonian, giving rise to a notion of geometric locality.

Note that when one introduces relativity or even a simple notion of geometric spacetime, the concept of quantum subsystems becomes more subtle than in pure quantum mechanics. For example, in the context of quantum reference frames, entanglement depends on the choice of reference frame. Since a natural decomposition into subsystems is based on entanglement, the notion of subsystems can itself be frame-dependent \cite{Giacomini2019}.

In this paper, we focus on the concept of subsystems rather than locality. We will not introduce any notion of geometric locality. Our approach to quantum mereology is restricted to non-relativistic quantum mechanics and is independent of the quantum state. Once again, we emphasize that it relies only on the spectrum of a system. This contrasts with previous approaches, such as refs \cite{Carroll2021, Summers2009, Zini2023, Cao2017} where the state plays an important role.

First, we will show that the decomposition into subsystems can be extracted from the spectrum. Then we detail a procedure to count the number of elementary subsystems from spectral properties. Finally we show how this can be used in practice to numerically find partitions that minimize entanglement growth.

\section{Partitioning the spectrum \label{sec:partitioning}}

\begin{figure}
\centering
\includegraphics[width=0.6\textwidth]{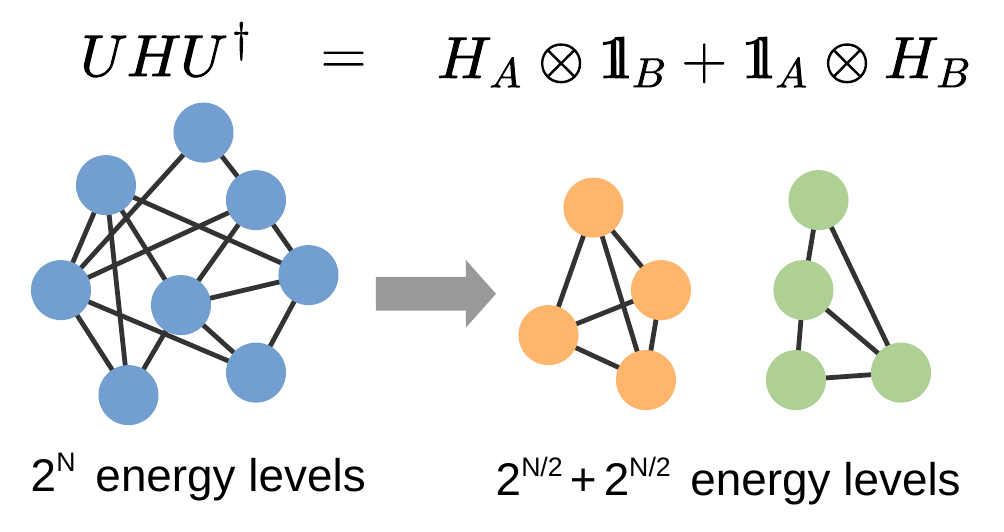}
\caption{A generic Hamiltonian $H$ comprises many body and all to all interactions between $N$ particles. After partitioning via a unitary $U$, the Hamiltonian $UHU^\dagger$ is represented in a new basis such that it consists in two sectors $H_A$ and $H_B$ each representing interactions between a subset of $N/2$ particles.}
\label{fig:drawing}
\end{figure}

The most basic mereology problem is partitioning: Given a quantum system, what is the best partition into weakly interacting subsystems? In this section we show that this problem can be solved in the energy eigenbasis alone, i.e. the question reduces to decomposing the spectrum of the initial system.

For simplicity, we will restrict ourselves to subsystems supported on qubits and to partitioning into two subsystems of the same dimension.
Starting with a generic Hamiltonian $H$ of dimension $D=2^N$, we want to find a tensor product structure that partitions $H$ into two weakly interacting sectors $A$ and $B$. This corresponds to finding a unitary $U$ and two Hamiltonians $H_A$ and $H_B$ of dimension $\sqrt D=2^\frac{N}{2}$ such that 
\begin{equation}
     UHU^\dagger =H_A\otimes \id_B +\id_A\otimes H_B+H_{int}\equiv H'
    \label{eq:partition}
\end{equation}
and $|H_{int}|$ is minimized.
Note that if this could be solved exactly, i.e if $H_{int}=0$ then $H'$ and $H$ would have the same spectrum. Thus we want to find $H_A$ and $H_B$ such that
\begin{equation}
    V\left(H_A\otimes \id_B +\id_A\otimes H_B\right) V^\dagger=E+\epsilon
\end{equation}
where $V$ diagonalizes $H'=H_A\otimes \id_B +\id_A\otimes H_B$, $E$ is the spectrum of the initial $H$ and $\epsilon$ is a diagonal error to be minimized. Note that $V$ can be decomposed as $V=V_A\otimes V_B$ where $V_A$ diagonalizes $H_A$ and $V_B$ diagonalizes $H_B$. Therefore, our problem is equivalent to finding two diagonal sets of energies $E_A$ and $E_B$ such that
\begin{align}
    E_A\otimes \id_B +\id_A\otimes E_B=E+\epsilon
    \label{eq:diag}.
\end{align}
We can consequently work in the particular case where $H$ is already diagonal.

\subsubsection*{Composition and convolution of the DOS}
Equation \ref{eq:diag} implies that if the partitioning is exact and $\epsilon=0$ then the spectrum of the whole system can be expressed as the sum of the spectra of the subsystems : $E_{nm}=A_n+B_m$.
This can be simply rephrased in terms of density of states (DOS) or its generating function: if $Z_A=\sum_ne^{-itA_n}$,  $Z_B=\sum_ne^{-itB_n}$ and $Z=\sum_{jk}e^{-itE_{jk}}$ then $Z=Z_AZ_B$. From the convolution theorem, the DOS of the whole system $\rho(E)=\int e^{-iEt}Z(t)dt$ is the convolution of $\rho_A$ and $\rho_B$. Therefore, the partitioning problem is equivalent to finding two DOS $\rho_A(E)$ and $\rho_B(E)$ such that their convolution $(\rho_A *\rho_B)(E)$ is the DOS of the system to be partitioned.

\subsubsection*{Gaussian density of states and central limit theorem}
Now using the central limit theorem, we can see that the composition of many generic systems will tend to have a Gaussian DOS. This is because the DOS of the composite system is effectively the convolution of the DOS of each individual system, and the central limit theorem implies that the convolution of many such independent distributions approaches a Gaussian distribution. Conversely, because the convolution of two Gaussians is another Gaussian, a Gaussian DOS can always be partitioned into two non-interacting systems with Gaussian DOS.

This suggests that the information about the subsystems is encoded in the perturbations away from the Gaussian spectrum. If a system has an exact Gaussian spectrum, then there is no preferred decomposition into elementary subsystems: the system can be partitioned into an infinite number of infinitely small non-interacting subsystems.

\section{Subsystems from the perturbations away from the Gaussian}

In this section, we will show how the number of elementary subsystems can be inferred just from the spectrum, or the density of states.
As suggested above, we will focus on the relation between subsystems and the perturbations away from the Gaussian spectrum.
Let's first prove that the DOS of a free model is Gaussian in the large $N$ limit, where $N$ is the number of independent subsystems (e.g. qubits). Such a model can be written as 
\begin{equation}
    H = \sum_i h_i \tau_i
\end{equation}
where $h_i$'s are real numbers and $\tau_i$'s are commuting Pauli strings: $[\tau_i,\tau_j]=\delta_{ij}\id$ and $\tr(\tau_i,\tau_j)=\delta_{ij}\tr\id$.
The density of states can be expressed as 
\begin{align}
    \rho(E) = \frac{1}{2\pi}\int_{-\infty}^\infty e^{-iEt}Z(H,t)dt,
    \label{rho}
\end{align}
and the partition function $Z(H,t)=\Tr(e^{itH})$ can be expanded in the moments $\mu_k(H)=\tr(H)^k$:
\begin{align}
    Z(H,t)&=\sum_{k=0}^\infty\frac{(it)^k}{k!}\mu_k\\
    &=\mu_0+it\mu_1-\frac{t^2}{2}\mu_2+\frac{it^3}{3!}\mu_3 ...
\end{align}
Therefore, the density of states is fully characterized by the moments $\mu_k(H)$. Moreover, the moments of the density of states are the moments of the Hamiltonian itself.
Computing the moments of the Hamiltonian is equivalent to a counting problem. For example, computing
\begin{equation}
\mu_4=\sum_{ijkl}h_ih_jh_kh_l\tr(\tau_i\tau_j\tau_k\tau_l)
\label{eq:mu4}
\end{equation} 
is equivalent to enumerating configurations $(i,j,k,l)$ so that $\tr(\tau_i\tau_j\tau_k\tau_l)$ is non zero. In general we have 
\begin{equation}
\label{eq:moments}
\mu_k=\sum_a \tr\prod^k_{l=1}h_{a_l}\tau_{a_l}
\end{equation}
where $a$ indexes all the possible contractions, e.g in the case $k=4$, $a=(i,j,k,l)$. 

In the case we have $M$ $\tau$'s that commute and the $h_i$ are drawn from a random variable with mean $0$ and width $\langle h^2\rangle$, then the moments are
\begin{equation}
    \mu_{2k} \approx (2k-1)!!\langle h^2\rangle ^k M^k 2^N
    \label{eq:moments2}
\end{equation} which are the moments of the Gaussian distribution \cite{Garcia2016, Garcia2017}. The term $(2k-1)!!$ comes from counting chord diagrams that pair the $\tau$'s two by two in the trace of eq. \eqref{eq:moments} such that the trace is non zero. 
For $2k=4$ there are 3 chord diagrams
\begin{equation}
\includegraphics[width=0.3\textwidth]{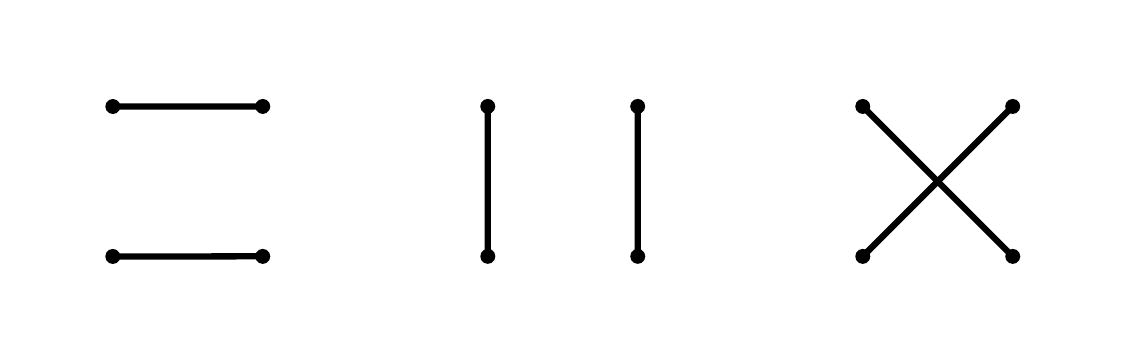}
\label{eq:chords4}
\end{equation}
corresponding to the terms $\tr(\tau_i\tau_i\tau_j\tau_j)$, $\tr(\tau_i\tau_j\tau_i\tau_j)$ and $\tr(\tau_i\tau_j\tau_j\tau_i)$.
For $2k=6$ there are 15 chord diagrams:
\begin{equation}
\includegraphics[width=0.35\textwidth]{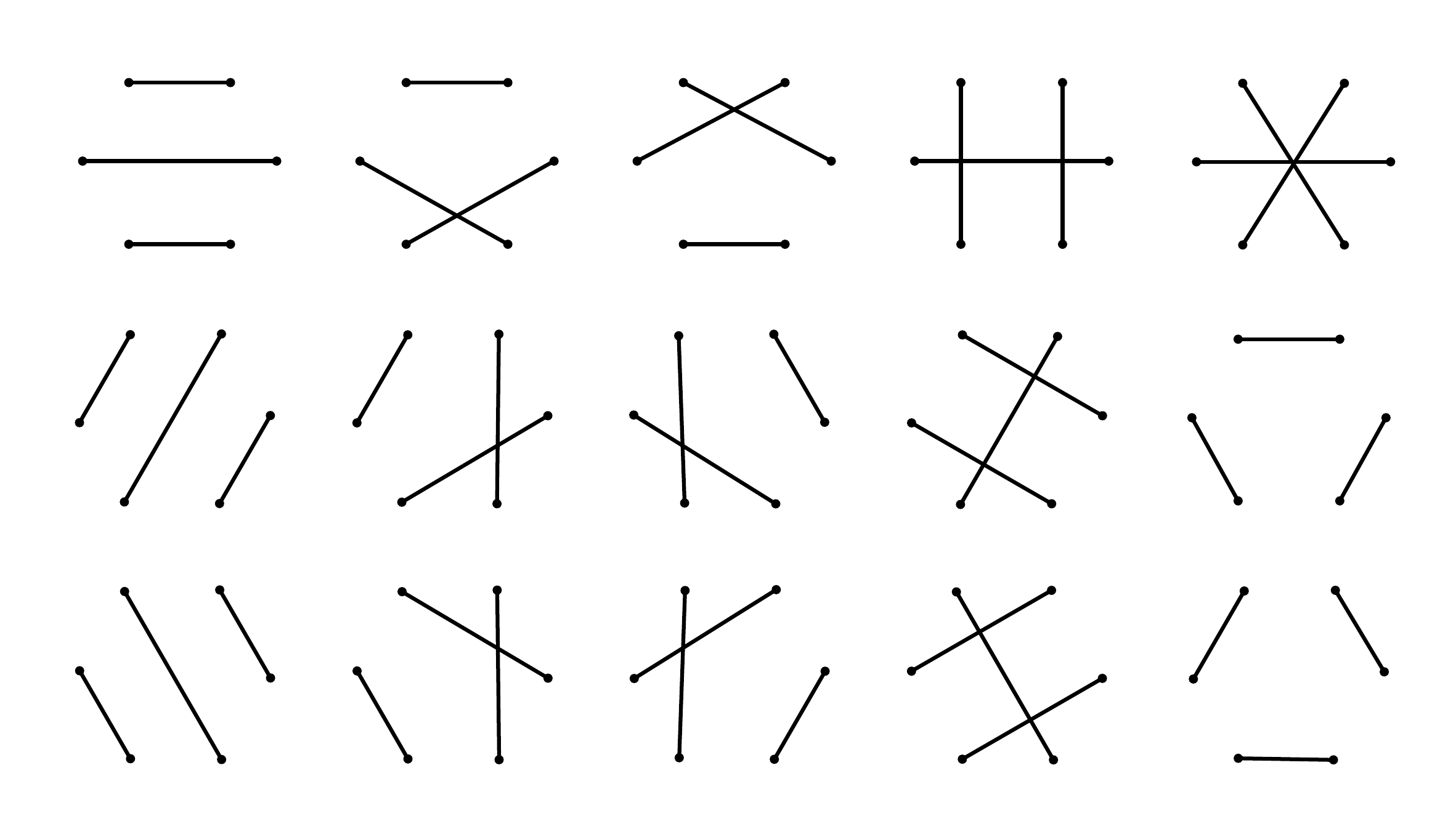}. 
\end{equation}
In these chord diagrams, each vertex represents the index of a $\tau$ and each line represent equality between indices. Each diagram represents a non zero term in equation \eqref{eq:moments}.
Then for every chord diagram, there are $M^k$ terms in the sum where $M$ is the number of $\tau$'s.

Consequently, as expected, the density of states of a free model is a Gaussian in the large $N$ limit. This is also true for other interacting local models where the probability that two randomly chosen strings $\tau_i$ and $\tau_j$ commute converges to one as the system size goes to infinity. For example, consider a 1D Ising system $H=\sum_i Z_i Z_{i+1}+\sum_i X_i$, then the Pauli strings are $Z_i Z_{i+1}$ and $X_i$ and the probability that two random strings do not commute drops like $\frac{1}{N}$. In the large $N$ limit, the density of states of all geometrically local models converges to a Gaussian. All the relevant thermodynamic information is therefore hidden in the perturbations away from the Gaussian. 

A simple way to study these perturbations is to look at the difference between the moments of a model and the moments of a Gaussian. 
First, define the normalized moments
\begin{equation}
    \mu_{2k}^N = \frac{\mu_{2k}}{\langle h^2\rangle ^k M^k 2^N}.
\end{equation}
such that they converge to the moments of the Gaussian $\mu_{2k}^\textup{gaussian}=(2k-1)!!$ in the large $N$ limit. In other words, they are the moments of $\rho(E)$ if its width is fixed to $1$.

\begin{figure}
\centering
\includegraphics[width=0.8\textwidth]{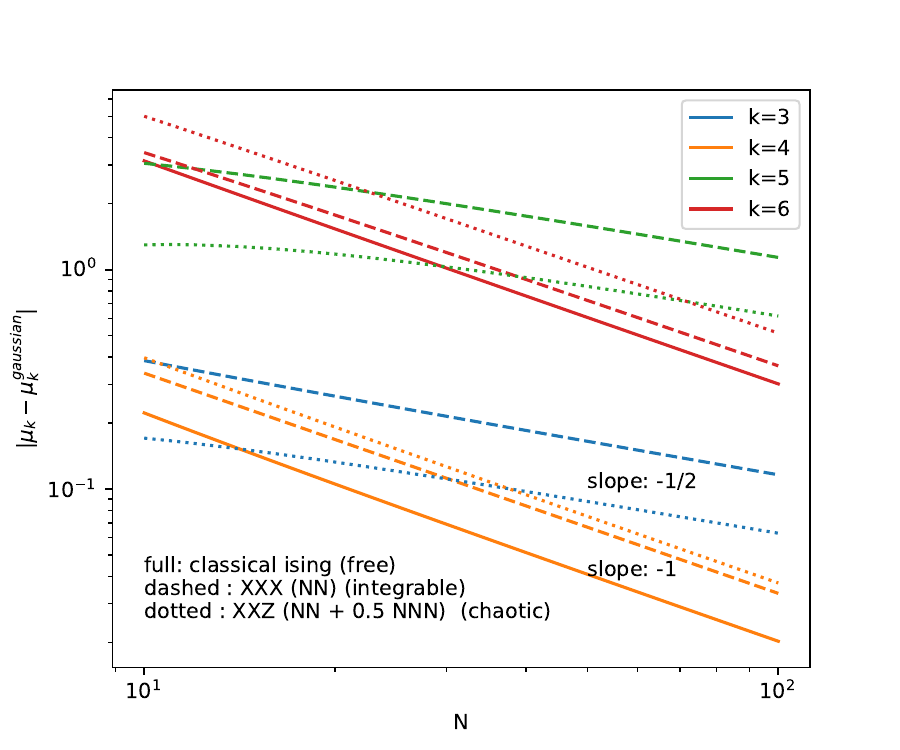}
\caption{Perturbations of the moments away from the Gaussian for a few 1D models. The full lines correspond to the classical Ising model with no field. The dashed lines are the XXX Heisenberg chain with nearest neighbor interactions. The dotted lines are the XXZ chain with next nearest neighbor interactions. For the even moments, the perturbations away from the gaussian decay like $\sim\frac{1}{N} $.}
\label{fig:1d}
\end{figure}

The perturbations away from the Gaussian can be written as $\Delta_{2k}=|\mu_{2k}^N-\mu_{2k}^\textup{gaussian}|$.
In figure \eqref{fig:1d} we show this quantify for $2k\in[3,6]$ for a few local models. We observe that $\Delta_4$ decreases like $\frac{1}{N}$. Let prove us this. In equation \eqref{eq:moments2} we made an approximation by only counting chords diagrams. Actually, each chord diagram also includes the fully connected diagram. In other words, the terms $\tr(\tau_i\tau_i\tau_j\tau_j)$, $\tr(\tau_i\tau_j\tau_i\tau_j)$ and $\tr(\tau_i\tau_j\tau_j\tau_i)$ include the term $\tr(\tau_i\tau_i\tau_i\tau_i)$. Therefore, $\tr(\tau_i\tau_i\tau_i\tau_i)$ needs to be subtracted twice from the fourth moment. A better approximation to the fourth moment in the case the $\tau$'s commute is
\begin{align}
\mu_{4}^\textup{} \approx 3\langle h^2\rangle ^2 M^2 2^N-2\langle h^2\rangle ^2 M 2^N
\end{align}
where the first term comes from diagrams \eqref{eq:chords4} and the second term comes from subtracting the fully connected diagram. The power of $M$ in the fully connected diagram is $1$ because there is only one sum over $i$. Now we can see that a better approximation to the fourth moment is:
\begin{equation}
    \mu_{4}^N \approx \mu_4^\textup{gaussian}-\frac{2}{M},
\end{equation}
meaning that the perturbation of the fourth moment away from the Gaussian decrease like $\frac{2}{M}$. In the case the model is geometrically local, the numbers of $\tau$ operators scales like $N$ and the perturbation decreases like $\frac{1}{N}$.

\begin{figure}
\centering
\includegraphics[width=0.6\textwidth]{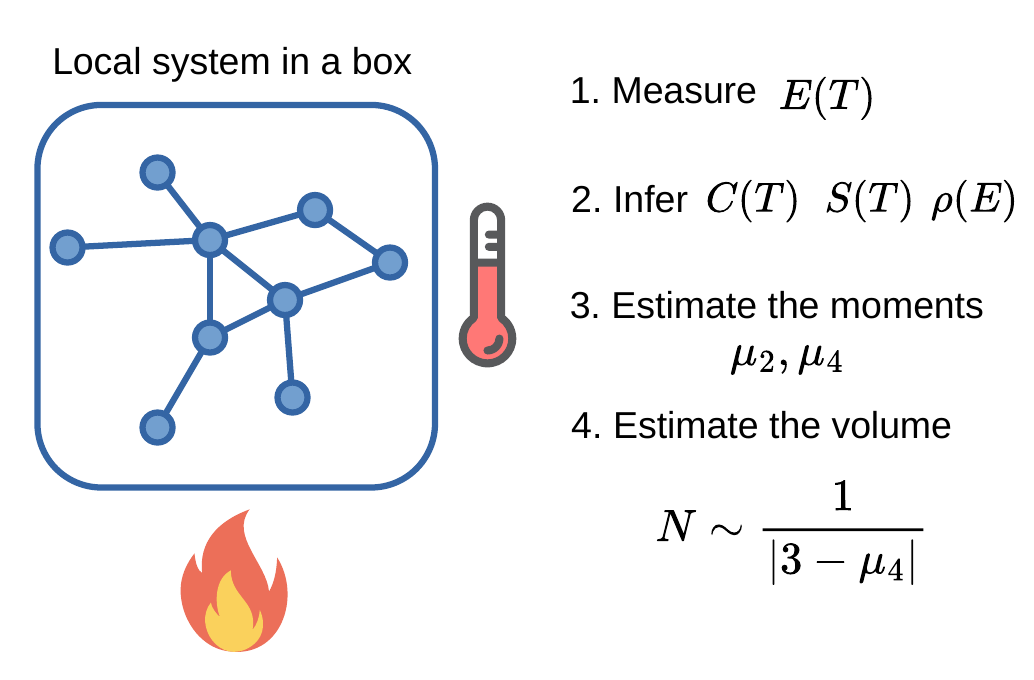}
\caption{Inferring the number of subsystems in a box by heating it. If one has access to $E(T)$ by heating the system, then one can infer the density of states $\rho(E)$ and estimate the number of subsystems.}
\label{fig:volume}
\end{figure}

This is an important result for quantum mereology because it suggests that one can read the number of subsystems from the density of states only. The procedure goes as follow : normalize the density of states such that its second moment is 1. Compute the fourth moment of the density of states, take its inverse, this is the volume in phase space occupied by the system i.e the number of elementary particles that constitute the system.

This suggests a more operational approach illustrated in figure \ref{fig:volume}. Consider you are given a black box containing a complex system and you need to count the number of subsystems without opening the box. By heating the box, you can measure $E(T)$, then compute the heat capacity $C=\frac{\partial E}{\partial T}$. Moreover, we have $\frac{C}{T}=\frac{\partial S}{\partial T}$ therefore can access $S(T)$ by integrating this expression. Finally, because we know $E(T)$, we can obtain the density of states $\rho(E)=e^{S(E)}$. From there we can measure the first moments and estimate the numbers of subsystems.

Note that our argument does not tell us anything about the dimension of the subsystems. What matters here is the algebraic structure of the $\tau$'s. We have chosen $\tau$'s to be Pauli strings for simplicity, but in general they can be any set of orthogonal operators. The subsystems can in principle be made arbitrary large by introducing symmetries in the $\tau$'s.

\section{Numerical applications}

\subsubsection*{Numerical partitioning}
\begin{figure}
\centering
\includegraphics[width=0.6\textwidth]{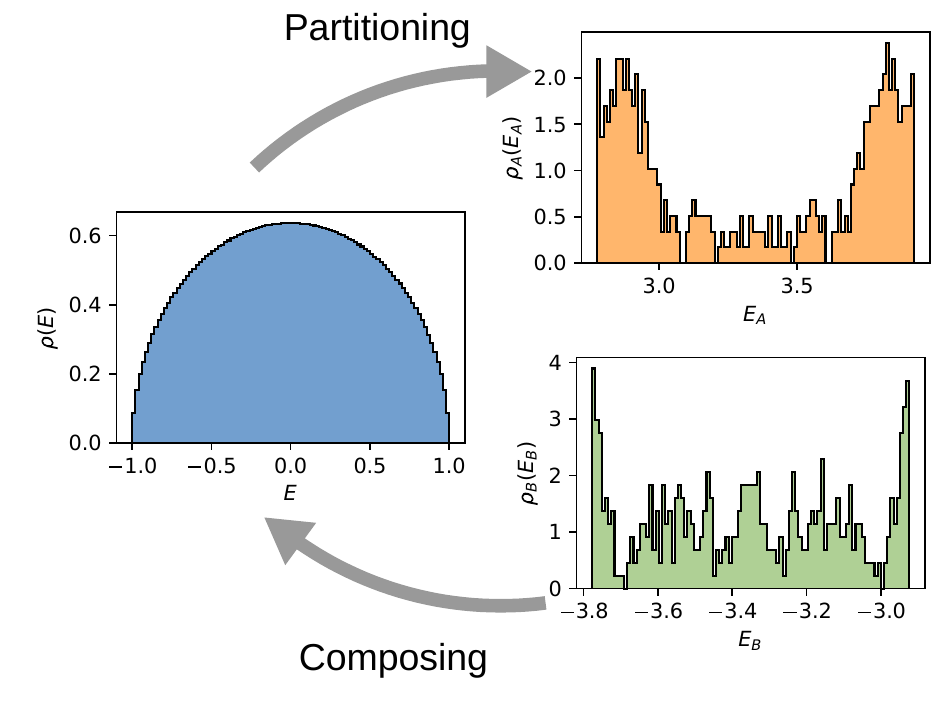}
\caption{Example of densities of states before and after partitioning (single realization). The left plot is the initial density of states $\rho$ with GOE statistics. The right plots show density of states $\rho_A$ and $\rho_B$ after partitioning of $\rho$. Note that the convolution $(\rho_A *\rho_B)(E)$ has to approximately yields $\rho(E)$. In order to recover the hard edges of the semicircle distribution $\rho_A$ and $\rho_B$ need to develop peaked edges.}
\label{fig:partitionAB}
\end{figure}

In this section we present a few numerical results that illustrate the idea that subsystems can be inferred from the spectrum itself. 
We attempt to partition random matrices into two weakly interacting sectors as illustrated in figure \eqref{fig:partitionAB}. We first generate spectra from the Gaussian Orthogonal Ensemble (GOE)  \cite{Wigner2023, Guhr1998, Atas2013}. 
As suggested by equation \eqref{eq:diag}, we can find a bipartition by minimizing the cost:
\begin{align}
    C &= \frac{1}{2^N}\sum_{ij}\left(E_{ij}-(A_i+B_j)\right)^2\label{cost}.
\end{align}
In practice, we do gradient descend
using the Broyden–Fletcher–Goldfarb–Shanno (BFGS) method \cite{BROYDEN1970,Fletcher1970,Goldfarb1970,Shanno1970,Fletcher2000}. Note that at every step, we need to sort the eigenvalues.
The method yields two spectra $A$ and $B$ and two permutations $P_1$ and $P_2$ such that 
\begin{align}
    P_1(E) = P_2(A \oplus B)
\end{align}
where $\oplus$ denotes the outer sum, $P_1$ sorts $E$ and $P_2$ sorts $A \oplus B$.
In terms of Hamiltonians, the method yields a unitary $U$ that encodes the change of basis giving the best partitioning of $H$. If $V$ diagonalizes $H$ into $P_1(E)$, then the partitioned Hamiltonian is $H'=UHU^\dagger$ with $U=P_2^{-1}V$.
Note that the best partitioning is not unique. Once a partitioning is found, any unitary acting on the partitions without scrambling subsystems $A$ and $B$, i.e. of the form $U_A\otimes U_B$, yields an equally good partitioning. The uniqueness of the solution beyond these classes of symmetries is an open question.

\begin{figure}
\centering
\includegraphics[width=0.6\textwidth]{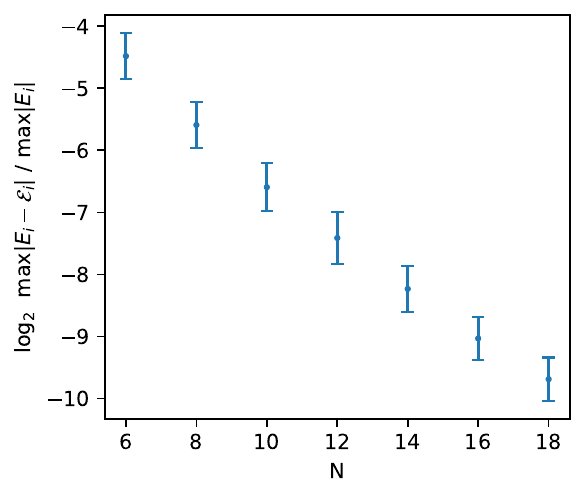}
\caption{Spectral norm $\log_{2}\frac{\max |E_i-\mathcal{E}_i|}{\max |E_i|}$ versus system size $N$ where $E$ is the spectrum of a $N$ qubit random GOE Hamiltonian and $\mathcal{E}$ is the spectrum of the partitioned Hamiltonian (equation \eqref{eq:partition}). Each data point is an average over 200 realizations and the error-bars show the standard deviation. The error is exponentially small in system size, meaning that large random matrices can be well partitioned into two non interacting sectors.}
\label{fig:cost}
\end{figure}

In practice, we generate random Hamiltonians from the Gaussian orthogonal ensemble (GOE) and attempt to partition them using our method. The GOE ensemble is a generic ensemble of random matrices that is commonly used to model chaotic quantum complex systems \cite{Wigner2023, Guhr1998, Atas2013}.
In figure \eqref{fig:cost} we show the spectral norm $\log_{2}\frac{\max |E_i-\mathcal{E}_i|}{\max |E_i|}$ where $E$ is the GOE spectrum and $\mathcal{E}$ is the spectrum of the partitioned Hamiltonian. The spectral norm is a good indicator of the convergence of the procedure since it quantifies the worst possible difference between the two spectra. In the GOE case, the spectral norm decreases exponentially with system size. This suggests that GOE matrices admit a natural partitioning into two very weakly interacting sectors.

\subsubsection*{Partitioning and slow entanglement growth}
If a spectrum can be partitioned successfully, it implies the existence of weakly interacting sectors that evolve independently over time without sharing information. A partition can also be interpreted as a preferred basis in which the entanglement growth of product states is slow. 
We quantify this by looking at the entropy $S(\Tr_B \oprod{\psi(t)}{\psi(t)})$ with $\psi(t)=e^{-itH'}\ket{\psi}_A\otimes \ket{\psi}_B$ where $H'=UHU^\dagger$ is the Hamiltonian in the partitioned basis and $\ket{\psi}_A$ and $\ket{\psi}_B$ are random pure states. Note that $H'$ is not $H_A\otimes \id_B +\id_A\otimes H_B$, it is $H$ in the basis that gives the best partition. In the case that the partition is perfect and there is no rest, then the entanglement entropy will be zero for all time, of course.

\begin{figure}
\centering
\includegraphics[width=0.6\textwidth]{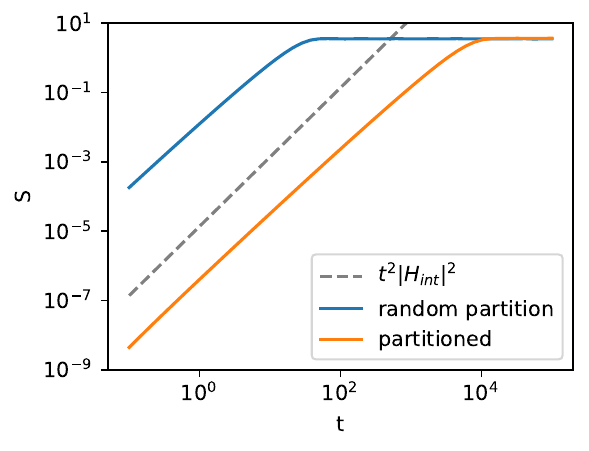}
\caption{Entanglement entropy growth of a product state under GOE Hamiltonian evolution for $N=10$. The blue line represent an arbitrary partition while the orange line shows the entanglement growth of the partitioned Hamiltonian. Note that both Hamiltonians have the same spectrum only the choice of the partitioning, i.e of the tensor structure differs. For reference, the grey line shows $t^2|H_{int}|^2$ where $H_{int}$ is the remaining interaction after partitioning. }
\label{fig:entropy_growth}
\end{figure}

In figure \eqref{fig:entropy_growth} we show that the growth of the entanglement entropy $S$ between subsystems $A$ and $B$ is $\sim 100$ times slower after partitioning than for a random partition for $N=10$. The entanglement growth is approximately quadratic in time. In geometrically local system, entanglement typically grows linearly, but here there is no notion of spatial geometry. The only objective of the optimization problem was to minimize $H_{int}$ the interaction between subsystems $A$ and $B$. Because $S$ is dimensionless and the entanglement growth should scale with $H_{int}$, we could expect $S$ to grow like $t^2|H_{int}|^2$. In practice, it grows even slower, as shown in figure \eqref{fig:entropy_growth}.

\section{Discussion}

We showed that quantum mereology can be restricted to reasoning on the spectrum : decomposing a quantum system into subsystems is equivalent to finding spectra that compose in a certain way. 

The limiting case is the Gaussian density of states: a Gaussian system can be partitioned indefinitely. No preferred subsystems emerge from a Gaussian distribution. In other words, a Gaussian spectrum does not encode any particular tensor structure.

Starting from this remark, we showed that information about the subsystems can be extracted from the spectrum by studying perturbations away from the Gaussian DOS. In the case of geometrically local models, the perturbation of the fourth moment scales as $\frac{1}{N}$. This provides an important mereological tool : counting the number of subsystems from the spectrum can be done by estimating the first moments of the density of states. 

Finally, as a practical example of quantum mereology, we developed a numerical procedure to partition quantum systems and applied this procedure to random matrices. 

Some more speculative ideas might arise from our results. Coming back to the idea of Democritus's atom : What happens if one tries to partition a quantum system again and again? Is it possible to partition a system forever? Is there a phenomenon that prevents us from partitioning a system indefinitely?

We argued that if the initial density of states is not exactly Gaussian, then we can count a finite number of elementary subsystems. This suggests that a system cannot be partitioned indefinitely. In this case, two different scales might emerge : 
\begin{itemize}
    \item At large scale it is easy to partition a system, systems are weakly interacting, observables supported on different subsystems almost commute (meaning that error $\epsilon$ in equation \eqref{eq:diag} is small), the ground state is localized and the physics of low energies and local observables can be well described in the diagonal basis,
    \item After multiple levels of partitioning, we reach a limit, subsystems cannot be partitioned anymore. This is where the quantum version of Democritus' atoms emerges and non-commutativity becomes relevant.
\end{itemize}

\begin{figure}
\centering
\includegraphics[width=0.6\textwidth]{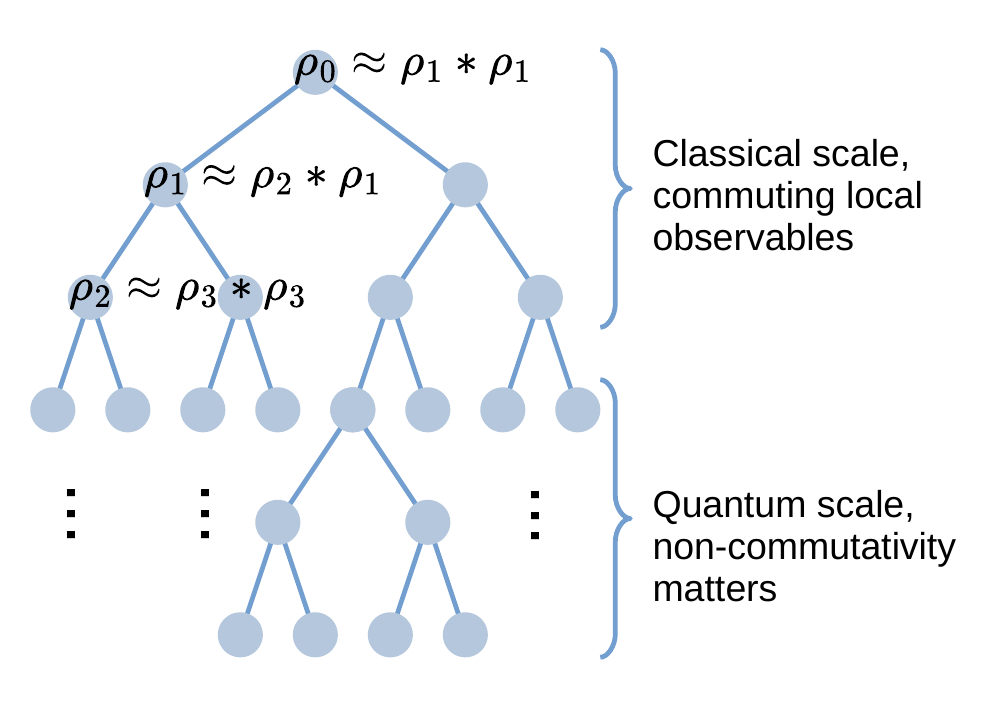}
\caption{Illustration of the recursive partitioning procedure. At each step, DOS of the subsystems can be approximated as the convolution of the DOS of lower level subsystems. If there is a limit to the partitioning depth, then two different scales emerge.}
\label{fig:partitiontree}
\end{figure}

If recursive partitioning fails at some scale, this is the scale at which the non-commuting properties of quantum mechanics appear. At the atom level, the best partitioning will be of the form 
\begin{align}
    H = H_A\otimes \id_B +\id_A\otimes H_B + H_{int}
\end{align}
where $H_{int}$ is non negligible. $H_{int}$ provides us with non-commutativity and with another smaller-scale preferred tensor structure. For example, it may be convenient to consider the change of basis $U=U_A\otimes U_B$ that conserves the subsystems $A$ and $B$ but factor $H_{int}$ into $UH_{int}U^\dagger=H_{int,A}\otimes H_{int,B}+H_r$ where the rest $H_r$ is minimized.

Finally, we suggest that the emergence of classicallity may be a tensor structure problem, as also mentioned in \cite{Adil2024, franzmann2024, Carroll2021}. A problem that consists in finding a tensor structure in which classical observables emerge and in which macroscopic subsystems behave classically.
When a system is partitioned into weakly interacting sectors, slow evolving local quantities arise, e.g. the total energy of each sector. Furthermore, any operator $O$ that commutes with $H_A$ the Hamiltonian of a subsystem, almost commutes with the total Hamiltonian $H = H_A\otimes \id_B +\id_A\otimes H_B + H_{int}$, the error is just $[O,H_{int}]$. Finding slowly evolving local macroscopic quantities is crucial to understand the emergence of classicallity from quantum many body systems.

\bmhead{Code availability}
A Python package for quantum mereology is available at \url{https://github.com/nicolasloizeau/quantum_mereology}.

\bmhead{Acknowledgements}
N.L. was partially supported by a research grant (42085) from Villum Fonden. We thank Soren Schlassa, Lennart Fernandes, and Flaviano Morone for their discussions.

\bibliography{sn-bibliography}

\end{document}